\documentstyle[prd,aps,doublespace,12pt]{revtex} 
\begin{document} 
%
%\begin{center}
\title{The Principle of Equivalence  
and Electro-magnetism.}
\author{G.R. Filewood}
\address{School of Physics,
University of Melbourne,
Parkville, Victoria 3052 Australia.}
\maketitle
\begin{abstract}
The problem of
unification of electro-magnetism
and gravitation in four dimensions;
 some new ideas involving 
the use of mixtures of commuting and
anti-commuting co-ordinates. Maxwell' s
equations are extracted in terms of curvature
of the anti-commuting part of space-time.
The profound difference in the coupling constants
of the two forces is interpreted in terms of
the degree of expansion of the two kinds of 
space-time with evolution of the universe.
Uncertainty in the quantum realm is interpreted
in terms of an unmeasurable  component of 
anti-commuting space-time.

Key words;
gravitation and electromagnetism, anti-commuting
co-ordinates,  Maxwell's equations.
\end{abstract}

\newpage
\section{Introduction}

The world of experience suggests to us
that space-time is a real 
continuum which may be represented
in terms of systems of commuting co-ordinates
composed of real numbers. It is thus natural that we should
extrapolate this experience to the world of
quantum objects and seek to describe them 
as embedded in a real commuting space-time.
However, the quantum domain is distinctly
different from our macroscopic experience so that one 
is led to question whether this extrapolation is
wholly appropriate.

In seeking an alternative to standard commuting
space-time one is led, by means of the
principle of relativity, to consider whether
there exists more general forms of co-ordinates
for which the laws of nature remain valid.
One favored means of further `generalizing'
space-time is to consider systems of commuting
co-ordinates of dimension higher than four
\cite{KK},\cite{Klein}.
However, another possibility which might be 
considered is that,
 in addition to systems of
commuting co-ordinates, one may have systems of
anti-commuting co-ordinates (or more generally
arbitrary mixtures of commuting and anti-commuting
co-ordinates). If anti-commuting co-ordinates exist in the
quantum domain then one may interpret their 
apparent absence in the macroscopic domain 
of experience in much the same way as one 
considers the absence of macroscopic spinorial
objects; over macroscopic distances the
anti-commuting co-ordinates effectively
`cancel-out'. 

We as  macroscopic observers cannot conceive of
a space-time of anti-commuting co-ordinates
for this is outside our domain of experience.
The world of macroscopic objects is in
a sense purely `bosonic' for the rotation
by $2\pi$ of any macroscopic object around
any (single) arbitrary axis brings it into self-congruence.
By contrast the wave function of 
a spinorial object changes  sign
 with a rotation by $2\pi$; an indication that, if an observer 
could become `spinorial', the observers experience of 
space-time would be very different to that normally
expected.
In standard theory spinors 
 such as electrons are represented by embedding them 
in a standard commuting L\"{o}rentz space-time.
This mathematical representation  works perfectly
well; although such success should not be taken
to indicate that such an embedding
 is necessarily both a complete and
correct description of space-time. 
 One hint that it may in fact be incomplete
comes from the failure of attempts to combine General Relativity
with quantum theory whereupon difficulties of
an extreme sort are encountered such as the Coleman-Mandula
(C.M.)
theorem {\cite{CM}} which forbids the (non-trivial)
union of compact and non-compact groups.

Is it possible that our extrapolation of commuting
space-time to the quantum domain, in spite of its
naturalness and success in the standard model, is
incorrect or incomplete?
 Is it possible that our interpretation of
the nature of spinorial matter is colored by our
preference for commuting space-time  for, at least
at the intuitive level, one might guess that 
a spinorial object such as an electron
will be  `bosonic'  in anti-commuting space-time? Is it
possible to avoid the C.M. theorem in
mixed commuting and anti-commuting co-ordinates?
(Supersymmetry, a symmetry connecting bosons and
fermions, is a known way of avoiding the C.M. theorem
 \cite{three}).

Putting aside for a moment the mathematical difficulties
involved in mixing commuting and anti-commuting
co-ordinates 
consider as an introduction to the premise of this paper
 the following.
Let us speculate that the universe began 
as a mixture of commuting and anti-commuting
co-ordinate space time in a roughly one-to-one
mixture. We might then further speculate that,
in keeping with the big-bang model, 
the commuting part of
space-time then expands  over time but the anti-commuting
part of space-time is far less prone to expansion
 because of a tendency of the
anti-commuting space-time to `cancel-out' over
macroscopic distances with inflation with the effective
fixed `range' of the anti-commuting co-ordinate
space-time  being
set by virtue of the quantisation of charge.

If the curvature of the commuting part of space-time then 
becomes the manifestation  of gravitation does there exist an
analogue of curvature of the anti-commuting part of 
space-time and what does it represent? It is the purpose of
this paper to explore these and related questions.
 Several things immediately
emerge to give us an indication of the possible answers.

The first is that, clearly, the curvature of the anti-commuting
space-time must relate to a massless gauge field in 
much the same way as gravitation will relate to a massless
gauge field. The reason is that the anti-commuting space-time
is not confined as such; its macroscopic effects are 
attenuated by cancelations rather than an abrupt finite
distance limit and thus its associated boson must be
massless for potential infinite range. There is only
one known 
(non-gravitation) 
candidate field quanta for this and it is the photon.

The second is that we may expect the coupling strength
ratio of the two massless gauge fields (one for
the commuting co-ordinates and one for the
anti-commuting co-ordinates) to evolve as a function of
effective relative expansion; beginning
at approximately the time of the big-bang with a ratio of 
order unity.
 This suggests that
the coupling constant of gravity is a dynamical
quantity related to the size and/or age of the universe
whereas the coupling constant of electro-magnetism
evolves as a function of the {\it{effective}} range
of the anti-commuting co-ordinates; the anti-commuting
space-time must have stretched only minimally
if at all
since the big-bang giving us an explanation for the
vast difference in the relative strength of the couplings
of gravitation and electromagnetism in the current epoch.

To form the appropriate non-inertial frame
to model the electro-magnetic field for the
electron  we must have co-ordinates capable of 
describing the space it is embedded in; since an
electron is a spinor the possibility that
a non-inertial anti-commuting frame may be
appropriate for this task warrants consideration.
 In this we can see an
analogue with general relativity. Inertial
co-ordinates are not adequate for a description
of the laws of nature for, from all possible
co-ordinate systems, they single out one 
preferred frame; the laws of nature should be 
valid in all frames. Likewise, the assumption
of purely commuting co-ordinates is unacceptable
since the laws of nature should be valid under
more general co-ordinates which  may
include some  component of anti-commuting 
co-ordinates or in general an arbitrary  mixture of 
commuting and anti-commuting co-ordinates.

In this way we hope to see electro-magnetism and gravitation
as two sides of one space-time. At a fundamental level
space-time will then be constructed of a continuum which
has both commuting and anti-commuting properties.
Curvature of the commuting aspect of space-time
is purely additive an leads to macroscopic fields
associated with macroscopic  objects such as stars.
 Curvature of the
anti-commuting aspect of space-time then must lead to
electric fields of particles and their spin-induced
magnetic moments. The inherently dual nature of 
anti-commuting space time leads to dual types
of fields; fields of attractive and repulsive
force which, over macroscopic distances
involving macroscopic matter, have
a tendency to cancel-out. The free electro-magnetic
field will then be interpreted as a propagating disturbance
in the  space-time vacuum due to temporary fluctuation away
from a net `zero' anti-commuting co-ordinates
in regions distant from charged sources.

To represent electro-magnetism as curvature we
need to extend the principle of equivalence.
 If we wish to describe
an object accelerating in space-time  we may 
choose to do so with respect to
 an inertial frame of reference.
The object in question will then not be traveling
with constant velocity with respect to such a
frame. However, we may alternatively choose
a non-inertial frame of reference - curved 
co-ordinates - with respect to which,
provided the form of the non-inertial  co-ordinates is
appropriately chosen, our accelerating
object describes a path of constant velocity.
This is just the statement that a non-inertial
frame is non-inertial i.e. accelerating.
It was the recognition of this simple fact
that led Einstein to G.R. for he realized
that, considering an object `at rest' on the surface of
the Earth as an accelerating object but one 
moving with constant velocity, one is immediately
let to deduce that the co-ordinates of the 
space are consequently not inertial. The
specific form of these non-inertial co-ordinates
then must describe the gravitational field
causing the `acceleration'. This is the principle of
equivalence; most succinctly put in terms of the
equivalence of inertial and gravitational mass
\cite{UFT},\cite{five}.

How might this principle be extended to
include electro-magnetism; a force which
bears much similarity to gravitation as
both are gauge fields  and both
are mediated by massless bosons (if one
accepts the reality of gravitons)?
Let us consider an electron's charge  as the analogue
of the mass which is the basis of the
principle of equivalence in gravitation
theory. One of the problems posed at the
beginning of this century by the 
atomic model was `why doesn't the electron,
since it is accelerated whilst it orbits
the nucleus, radiate energy  and spiral
into the positively charged nucleus?' The
answer given was that the energy of the
electron, and any energy it radiated, was quantized
and could only come in quantum `jumps' not a continuum.

However there is another way of looking at this
problem in terms of the above postulate. If 
we consider a charged object such as an electron
moving with constant velocity in the vacuum
we may think of the surrounding electric field
of the electron as approximately a symmetric 
(in terms of the direction of motion) finite radius 
of anti-commuting space-time (by `finite' it is
implied that the field is $\approx0$ at macroscopic
distances). The effective radius of the field,
which we may define at some arbitrary strength
cut-off,
may be taken as  a constant for an inertial observer
independent of where the cut-off is defined. However if
the charge is {\it{accelerating}} 
with respect to an inertial observer
an asymmetry
will develop in the field between the forward  direction (of 
motion) and backward direction anti-commuting space-time
 and this may be interpreted in terms of generated 
waves in the anti-commuting space-time;
photons in this schema. That an electron in a 
hydrogen atom does not radiate energy then implies
that the electron is traveling only in a straight
line with constant velocity in terms of anti-commuting 
space-time and so does not radiate energy; the
anti-commuting 
space must then be curved. The quantized nature of the
possible paths for the electron then arises because
the analogue of mass - in this case the charge of
the particle - comes in quantum units not in
a continuum (the possible masses of macroscopic
objects may be regarded as a continuum).

Thus we shall seek to recast general relativity 
in these more generalized co-ordinates. In
so doing we shall  seek to  extend the principle
of equivalence to encompass the concept of
curvature in anti-commuting space-time
as a representation of the electro-magnetic field.
This principle will be  expressed as  the
vanishing of the covariant derivative of the
metric; an expression that a frame exists
{\it{locally}} in anti-commuting co-ordinates
in which the field can be `transformed away';
the analogue of a free-fall frame.

As an introduction consider  the (L\"{o}rentz covariant)
Gupta-Bleuler formalism for photon quantization
in which we have the following commutator;
\[
[a^{\alpha}(k),a^{\dagger\beta}(k')]_{-}
=
-g^{\alpha\beta}_{s}\delta^{3}(k-k')
\]
where $g^{\alpha\beta}_{s}$ is the standard
symmetric metric diag.$(+,-,-,-)$ with the suffix
`s' indicating symmetric. Using
$g^s_{\alpha\beta}g_s^{\beta\alpha}=+4$ this is  rewritten;

\begin{equation}
[a^{\alpha}(k),a^{\dagger}_{\;\alpha}(k')]_{-}
=
-4\delta^{3}(k-k')
\label{a1}
\end{equation}
However if $g^{\alpha\beta}{\equiv}g_a^{\alpha\beta}$
 is purely {\it{antisymmetric}} and if it
also has the appropriate mathematical properties to
raise and lower the indices of the creation and 
annihilation operators $a^{\dagger}$ and $a$
then its substitution into the commutator produces;
\begin{eqnarray}
-\bar{g}^{a}_{\alpha\beta}
[a^{\alpha}(k),a^{\dagger\beta}(k')]_{-}
&=&
a_{\beta}(k).a^{\dagger\beta}(k')
+a^{\dagger}_{\alpha}(k').a^{\alpha}(k)
\\
\nonumber
&=&
\{a^{\alpha}(k),a^{\dagger}_{\alpha}(k')\}_{+}
\\
\nonumber
&=&
c.\delta^{3}(k-k')
\label{a2}
\end{eqnarray}
where $\bar{g}^{a}_{\alpha\beta}$
is the complex-conjugate transpose or `dual'
metric and for some constant 
$c=\bar{g}^{a}_{\alpha\beta}{g}_{a}^{\beta\alpha}$
 (provided $c\neq0$)
a commutator has been converted 
to an anti-commutator; i.e. the interchange of a 
symmetric and anti-symmetric metric has implied
an interchange of particle description from
spin 1 to  ${1\over2}$; this implies 
 that the particle statistics have changed! 
This is of course inconsistent with observation.
 However, if we
add, in addition to the substitution of
an anti-symmetric metric anti-commuting
co-ordinates then we may suspect that  the 
proper statistics will be restored. Thus we may suspect that
general  admixtures of symmetric
and anti-symmetric metric, if appropriately
combined with mixtures of commuting and anti-commuting
co-ordinates,  is covertly `supersymmetric'
if the laws of nature are invariant with respect 
to whatever arbitrary
 admixture is
chosen. This approach  however differs
fundamentally from conventional supersymmetry in that
no super-partner particles ever arise. 

This simple exercise gives some hope 
that by modifying the structure of space-time
a property somewhat analogous to supersymmetry,
with the concomitant hope of avoidance of the
C.M. theorem, might be embedded into the
structure of space-time itself. In the next sections
 mathematical machinery is developed to 
give effect
to such a structure.

\section{Metric  and co-ordinates.}

Our first task it find the form of the
metric and the co-ordinates appropriate
for an `inertial frame' in  anti-commuting 
space-time. We must then seek to extend 
the principle of equivalence to non-inertial
anti-commuting co-ordinate systems.
The introduction of anti-commuting
co-ordinates means that the metric
will also have to be modified to
include an anti-symmetric component;
this is so that the invariant length
\[
ds^2={\delta}x_a^{\alpha}g^s_{\alpha\beta}{\delta}x_a^{\beta}
+
{\delta}x_a^{\alpha}g^a_{\alpha\beta}{\delta}x_a^{\beta}
\]
is well defined in anti-commuting
co-ordinates. Here  s = symmetric and 
a = anti-symmetric (so that $x_{a}^{\alpha}x_{a}^{\beta}
=-x_{a}^{\beta}x_{a}^{\alpha}\;;\;{\forall}\alpha\neq\beta$);
 the co-ordinates only commute if they
have the same index (this is required so that
$x_a^{\alpha}x^a_{\alpha}\neq0$). If the combined
metric $g_{\alpha\beta}=g^s_{\alpha\beta}
+g^a_{\alpha\beta}$ is an invariant tensor
in the theory then $ds^2$ is also an
invariant.

Consider the following metric which is a 16x16
matrix but has only four space-time indices;
\begin{equation}
{1\over2}
\left({\begin{array}{cccc}
-I_4&{i}\sigma_{01}&{i}\sigma_{02}&
{i}\sigma_{03}\\{i}\sigma_{10}&+I_4&
{i}\sigma_{12}&{i}\sigma_{13}\\
{i}\sigma_{20}&{i}\sigma_{21}&+I_4&
{i}\sigma_{23}\\{i}\sigma_{30}&
{i}\sigma_{31}&{i}\sigma_{32}&+I_4\
\end{array}}\right)
\equiv
-{1\over2}I_4.\eta_{\alpha\beta}+
{i\over2}\sigma_{\alpha\beta}
\label{sigmametric}
\end{equation}
\\
where $\eta_{\alpha\beta}=$diag.$(+,-,-,-)_{\alpha\beta}$
and
$\sigma^{\alpha\beta}=
{i\over2}\left[\gamma^{\alpha},\gamma^{\beta}
\right]$. As a consequence of the 
choice of  $\eta_{\alpha\beta}$ the gamma
matrices follow the form $\gamma^{{\dagger}0}
=\gamma_0=\gamma^0$ and 
$\gamma^{{\dagger}i}=\gamma_{i}=-\gamma^{i}$.
 This metric has the following
closure property;

\begin{eqnarray}
g_{\alpha\phi}\bar{g}^\phi_{\,\,\,\beta}
&=&{1\over4}\left(-I_4.\eta_{\alpha\phi}
+i\sigma_{\alpha\phi}\right)\left(
-I_4.\eta^\phi_{\,\,\,\beta}-i\sigma^{\phi}
_{\,\,\,\,\beta}\right)\nonumber\\&=&
{1\over2}
\left(-I_4.\eta_{\alpha\beta}
+i\sigma_{\alpha\beta}\right)
=g_{\alpha\beta}
\label{consistencyconstraint}
\end{eqnarray}
where I have used $\gamma^0\gamma_0=
\gamma^1\gamma_1=
\gamma^2\gamma_2=
\gamma^3\gamma_3=I_4$
\footnote
{most Q.F.T. textbooks contain the relation
$\gamma^\alpha\gamma_\alpha=4$
with a summation convention on $\alpha$
but strictly speaking the gamma matrices
square up to a 4x4 identity
matrix not a scalar in which case
$\gamma^\alpha\gamma_\alpha=4I_4$.} and
 $\bar{g}$ is the  `dual' metric
viz; 
\begin{equation}
\left(\sigma_{\alpha\beta}
\right)^\dagger=
{-i\over2}\left[\gamma_\alpha,
\gamma_\beta\right]^\dagger=
\sigma^{\alpha\beta}
\label{phase}
\end{equation}
viz $\gamma^{{\dagger}0}=\gamma_0=\gamma^0$
and $\gamma^{{\dagger}i}=\gamma_i=-\gamma^i$
with $\eta_{\alpha\beta}=\eta^{\alpha\beta}$ a
purely real diagonal matrix so that; 
\begin{equation}
\left
(g_{\alpha\beta}\right)^\dagger
=\bar{g}^{\alpha\beta}=
{1\over2}\left(-I_4.\eta^
{\alpha\beta}-i\sigma^{\alpha
\beta}\right)
=g^{\beta\alpha}
\label{inverse}
\end{equation}
The `bar' operation is represented by taking the
complex-conjugate transpose and raising (lowering)
all indices.
 Using a complex metric such as metric.(\ref{sigmametric})
requires generalization 
 of the L\"{o}rentz transformation
\footnote{see appendix A.}. Under the 
modified L\"{o}rentz transformation the metric transforms
as an invariant  tensor as is required for a metric;
\begin{equation}
g_{\mu^{'}\nu^{'}}
=
\bar{\Lambda}_{\mu^{'}}^{\;\;\mu}
g_{\mu\nu}{\Lambda}_{\nu^{'}}^{\;\;\nu}
\end{equation}

There is a further complication concerning the 
use of metric.(\ref{sigmametric}) as a space-time
metric. Consider the following transformation of
the co-ordinates under the modified L\"{o}rentz
transformation given in the appendix;
\begin{equation}
\left(x^{\alpha'}\right)^{\dagger}
=
\left(
\bar{\Lambda}
^{\alpha'}_{\;\;\;\alpha}x^\alpha
\right)^\dagger
=
x^{\dagger\alpha}\bar{\Lambda}^{\dagger\alpha'}
_{\;\;\;\;\alpha}
=x^{\dagger\alpha}\bar{\Lambda}_{\;\;\;\alpha'}
^{\alpha}
\end{equation}
where the last step follows from the property 
of the metric under complex-conjugate transpose
eq.(\ref{inverse}). Thus we require that the
co-ordinates have the  property
$
x^{\dagger\alpha}=x_\alpha$
i.e. the co-ordinates must act under 
complex-conjugation like the gamma matrices.

Now this can be achieved because of the
following  property of metric (\ref{sigmametric});
\begin{equation}
\bar{g}_{\alpha\beta}{\gamma^{\beta}\over2}
={\gamma_{\alpha}\over2}
\;\;\;\;\mbox{and}\;\;\;\;
{g}_{\alpha\beta}{\gamma^{\dagger\beta}\over2}
={\gamma_{\alpha}\over2}
\end{equation}
which,
for the chosen representation $\gamma^{\dagger0}
=\gamma_0,\;\;\gamma^{{\dagger}i}=\gamma_i$,
 is a property one normally only expects to
find in the purely symmetric diagonal metric $\eta=(+,-,-,-)$.
 In order that 
the constraint $x^{\dagger\alpha}=x_\alpha$ is 
satisfied it is necessary to couple each space-time
co-ordinate to a gamma  matrix so that, for example,
the scalar
$x^0$ gets multiplied into each entry of a
$\gamma^0$ matrix and becomes $
{1\over2}x^0.\gamma^0{\equiv}x^{\not\,0}$ and
${1\over2}x^1.\gamma^1{\equiv}x^{\not\,1}$ etc. From now on
slashed  $x^{\not\alpha}$ means space-time co-ordinates coupled
to the corresponding gamma matrix; i.e. each 
co-ordinate  is now represented by a 4x4 matrix.
 To indicate that 
the corresponding 16x16 metric is intended to contract
against these modified co-ordinates the metric
indices are also labeled with a slash i.e.
\[
g_{\not\alpha\not\beta}\equiv
{-I_4\over2}\eta_{\alpha\beta}
+{i\over2}\sigma_{\alpha\beta}.\]
It is important to realize that the information
as to  whether a co-ordinate
index is `upstairs' or `downstairs' is now
carried on the gamma matrix index {\it{not}} on
the co-ordinate scalar so for the scalars
$x^0=x_0, x^1=x_1$ etc. whilst for the
matrices $x^{\not\,0}=x_{\not\,0}
,x^{\not\,1}=-x_{\not\,1}$ etc.

Note that the invariant length $ds^2$ is now
positive definite (but only if translated
into terms of commuting co-ordinates);
 i.e. related to a compact group;
\begin{eqnarray}
ds^2&=&dx^{\not\alpha}\bar{g}_{\not\alpha\not\beta}
dx^{\not\beta}
\\
\nonumber
&=&{1\over4}(dx^0\gamma^0\gamma_0dx_0+
dx^1\gamma^1\gamma_1dx_1
+dx^2\gamma^2\gamma_2dx_2+
dx^3\gamma^3\gamma_3dx_3)
\\
\nonumber
&=&dx^0dx_0+dx^idx_i
\end{eqnarray}
where the last step follows because ds is strictly a
scalar; that is, the trace must be taken over the 4x4
identity matrices which result from the product of each 
pair of gamma matrices i.e. $Tr\gamma^0\gamma_0=TrI_4=4$ etc.
But note that the L\"{o}rentz structure is still manifest if
we keep the anti-commuting structure;
\begin{eqnarray}
ds^2=dx^{\not\alpha}dx_{\not\alpha}&=&
(dx^{\not\,{0}})^2-(dx^{\not\,{1}})^2
-(dx^{\not\,{2}})^2-(dx^{\not\,{3}})^2
\\
\nonumber
&=&
(dx_{\not\,{0}})^2-(dx_{\not\,{1}})^2
-(dx_{\not\,{2}})^2-(dx_{\not\,{3}})^2
\end{eqnarray}
 In order that 
the derivative has the appropriate 
properties with respect to the metric
the derivative must also be coupled to
the gamma matrices; a form appropriate for
a quantum mechanical operator formalism is used;
$\partial_\mu\rightarrow{-i\over2}\partial_\mu
\gamma_{\mu}\equiv{\partial}_{\not\mu}$;
(note that this symbol is not the same as
$\not\!\partial=\partial_{\mu}\gamma^{\mu}$
as in the former there is no sum over $\mu$;
the summation convention will be followed  only for summation between
one `up-stairs' and one `down-stairs' index; two identical
indices either both `up-stairs' or both `down-stairs' are not
summed). Note that $\partial_{\not\mu}^{\dagger}=
-\partial^{\not\mu}$ so that $\overline{\partial_{\not\mu}}
=-\partial_{\not\mu}$.
The metric also has the 
following property;
\begin{equation}
g_{\not\alpha\not\epsilon}g^{\not\epsilon}_{\;\;\not\beta}
=I_4\eta_{\alpha\beta}
\end{equation}
that is, that in some sense the metric is the
square root of the normal diagonal L\"{o}rentz
metric. The inverse is simply;
\begin{equation}
g_{\not\alpha\not\epsilon}g^{\not\epsilon\not\beta}
=I_4\eta_{\alpha}^{\;\;\;\beta}
=\delta_{\not\alpha}^{\;\;\;\not\beta}
\end{equation}

A group with a positive definite
metric is compact (here it has an invariant length S 
equivalent to $O_4$). The unusual feature of
metric (\ref{sigmametric}) is that mathematical
structures associated with the
orthochronous L\"{o}rentz group, in this case
the fundamental tensor  and the 
spinor rep, are summed to generate a compact group.
But this results only obliquely by reference to a translated
co-ordinate system; in the co-ordinate system 
appropriate for the metric, an anti-commuting 
co-ordinate system, the topology is 
non-compact!

\section{Covariant Derivative and vanishing non-metricity.}

The next step is to generalize the metric
(\ref{sigmametric})
to non-inertial frames.
In order to define covariant differentiation
in anti-commuting co-ordinates special emphasis is 
placed on the vanishing of the covariant
derivative of the metric; this is the mathematical
expression of the principle of equivalence since 
it means that a co-ordinate system can be found
in which the force in question - in G.R. this
is gravitation but in the theory being developed
here it will include electro-magnetism - can be
`transformed away'; a `free-fall' frame
exists {\it{locally}} in which the force is `absent'.
 (The electro-magnetic field cannot
be `transformed away' by a L\"{o}rentz transformation
in commuting co-ordinates).

Firstly form the generalized connection;
\begin{equation}
\Gamma_{\not\alpha\not\beta}^{\not\rho}={1\over2}
\left(g_{\not\,\epsilon\not\alpha,\not\beta}+g_
{\not\beta\not\,\epsilon,\not\alpha}
-g_{\not\alpha\not\beta,\not\,\epsilon}\right)
\bar{g}^{\not\,\epsilon\not\rho}
\label{zot}
\end{equation}
where the order of indices in eq.(\ref{zot}) is
to be maintained rigorously. The dual is;
\begin{equation}
\bar{\Gamma}_{\not\alpha\not\beta}^{\not\rho}={1\over2}
\bar{g}^{\not\rho\not\,\epsilon}
\left(\overline{{g}_{\not\,\epsilon\not\alpha\,,\,\not\beta}}+
\overline{{g}_
{\not\beta\not\epsilon\,,\,\not\alpha}}
-\overline{{g}_{\not\alpha\not\beta\,,\,\not\,\epsilon}}\right)
\end{equation}
and again the order of indices is to be maintained. Note that the
dual involves the derivative. The derivative $,\not\!\!\gamma$
is taken to be acting to the left. The bar operation is
equal to hermitian conjugation followed by lowering (or
raising) all indices;
\begin{equation}
(\overline{g_{\not\alpha\not\beta\;,\;\not\gamma}})
=
({g}_{\not\alpha{^{\dagger}}\not\beta
{^{\dagger}}}\stackrel{\leftharpoonup}
{\partial_{\not\gamma^{\dagger}}})^{\dagger}=
-\stackrel{\rightharpoonup}
{\partial_{\not\gamma}}\bar{g}_{\not\alpha\not\beta}
=
-\stackrel{\rightharpoonup}
{\partial_{\not\gamma}}{g}_{\not\beta\not\alpha}
\end{equation}
The
definition of the dual connection is not required to define 
the covariant derivative of the metric  but is
needed to prove theorems for more general forms
involving derivatives of vector fields.
As we shall see, we are interested only in the situation that the
derivative of the symmetric part of the metric
vanishes. In that circumstance;\[
g_{\not\alpha\not\beta\;;\;\not\gamma}
=
\bar{g}_{\not\beta\not\alpha\;;\;\not\gamma}
=
-{g}_{\not\beta\not\alpha\;;\;\not\gamma}\]
and the order of indices in the differentiation
is important since they do not necessarily commute. In
keeping with the form of the notation the covariant
derivative index is understood to be acting on the index
to its immediate left. Thus the covariant derivative
of the metric is written as;
\begin{eqnarray}
g_{\not\alpha\not\beta\;;\;\not\gamma}
=
g_{\not\alpha(\not\beta\;;\;\not\gamma)}
+\bar{g}_{\not\beta(\not\alpha\;;\;\not\gamma)}
&=&
g_{\not\alpha(\not\beta\;,\;\not\gamma)}
+\bar{g}_{\not\beta(\not\alpha\;,\;\not\gamma)}
-\Gamma_{\not\beta\not\gamma}^{\not\rho}
\;g_{\not\alpha\not\rho}
-\bar{\Gamma}_{\not\alpha\not\gamma}^{\not\rho}
\;\bar{g}_{\not\beta\not\rho}
\label{dual}
\end{eqnarray}
or alternatively, in the circumstance that the derivative of 
the symmetric part of the metric vanishes (which is
the situation under consideration here) 
the connection is anti-symmetric in its
lower two indices and we have;
\begin{eqnarray}
g_{\not\alpha\not\beta\;;\;\not\gamma}
=
g_{\not\alpha(\not\beta\;;\;\not\gamma)}
-{g}_{\not\beta(\not\alpha\;;\;\not\gamma)}
&=&
g_{\not\alpha(\not\beta\;,\;\not\gamma)}
-{g}_{\not\beta(\not\alpha\;,\;\not\gamma)}
-\Gamma_{\not\beta\not\gamma}^{\not\rho}
\;g_{\not\alpha\not\rho}
+{\Gamma}_{\not\alpha\not\gamma}^{\not\rho}
\;{g}_{\not\beta\not\rho}
\nonumber\\
&=&
g_{\not\alpha\not\beta\;,\;\not\gamma}
-\Gamma_{\not\beta\not\gamma}^{\not\rho}
\;g_{\not\alpha\not\rho}
-{\Gamma}_{\not\gamma\not\alpha}^{\not\rho}
\;\bar{g}_{\not\rho\not\beta}
\nonumber\\
&=&0
\label{covariantderivative}
\end{eqnarray}
where the last line follows directly from the definition
of the connection and the inverse property of the metric.
Using (\ref{dual}) and (\ref{covariantderivative})
we have;\[
\bar{\Gamma}_{\not\alpha\not\gamma}^{\not\rho}
\;{g}_{\not\rho\not\beta}
={\Gamma}_{\not\gamma\not\alpha}^{\not\rho}
\;\bar{g}_{\not\rho\not\beta}\]
which is a relation derived from a
general covariant derivative and should
therefore hold for any rank two tensor in the
theory if the covariant derivative of the
metric is to vanish; i.e.,
\begin{equation}
\bar{\Gamma}_{\not\alpha\not\gamma}^{\not\rho}
\;{u}_{\not\rho\not\beta}
={\Gamma}_{\not\gamma\not\alpha}^{\not\rho}
\;\overline{{u}_{\not\rho\not\beta}}
\label{conn}
\end{equation}
for any rank two tensor u. This relation can be used
to generalize the covariant derivative to
higher rank tensors. In particular
if the vector field is real 
($\overline{A_{\not\mu}}=A_{\not\mu}$) 
and satisfies the L\"{o}rentz
condition
we have;
\begin{equation}{A_{\not\mu\,,\,
\not\nu}}=-g^s_{\not\mu\not\nu}
A_{\;\;\;,\not\rho}
^{\not\rho}-{\stackrel{\rightharpoonup}{\partial_{\not\nu}}}
A_{\not\mu}
=-{\stackrel{\rightharpoonup}{\partial_{\not\nu}}}
A_{\not\mu}
=
\overline{A_{\not\mu\,,\,\not\nu}}
\label{Lorentz1}
\end{equation}
and by the definition of the bar operation
for a purely real field;
\begin{equation}
{A_{\not\mu\,;\,\not\nu}}
=
{A_{\not\mu\,,\,\not\nu}}
-{\Gamma}_{\not\mu\not\nu}^{\not\rho}
{A_{\not\rho}}=
\overline{A_{\not\mu\,,\,\not\nu}}
-\bar{\Gamma}_{\not\nu\not\mu}^{\not\rho}
\overline{A_{\not\rho}}
=
\overline{A_{\not\mu\,;\,\not\nu}}
\label{can}
\end{equation}
Notice that the bar operation here involves the derivative 
and is the source of the 
odd permutation of un-contracted indices
in the barred connection on the R.H.S. of eq(\ref{can}).
We can now define the general iterated  covariant derivative
 for a real vector field $A_{\not\alpha}$
satisfying the
L\"{o}rentz gauge 
(viz.  eqs.(\ref{dual}), (\ref{conn}) and (\ref{can}));
\begin{eqnarray}
A_{\not{\alpha}\;;\;\not\beta\;;\;\not\gamma}
&=&
A_{\not\alpha\;;\;\not\beta\;,\;\not\gamma}
-{\Gamma}^{\not\phi}_{\not\gamma\not\alpha}
{A_{\not\phi\;;\;\not\beta}}
-\Gamma^{\not\phi}_{\not\beta\not\gamma}A_{\not\alpha\;;\;\not\phi}
\end{eqnarray}
 Note that in
forming the covariant derivative an
even permutation of the indices $\not{\!\alpha}
\not\!\beta$ and $\not{\!\gamma}$ is maintained on the R.H.S.
of eq.(\ref{covariantderivative})
with respect to the order they appear  in the
metric on the L.H.S. (either g or $\bar{g}$ as the case may be)
as we would do for spinor indices.

\section{Curvature and covariant differentiation
in anti-commuting co-ordinates}

In the circumstance that the derivatives of the
symmetric part of the metric vanish (gravity ignored;
a free-fall frame) the connection is purely anti-symmetric
in its lower indices (this nominally indicates a 
torsion tensor but in the present case, because
the co-ordinates anti-commute, the physics represented
will not in fact be torsion).
In that context
consider the {\it{sum}} of the following two-fold 
covariant derivatives of a real vector field; noting that
in re-ordering the indices 
when expanding out in terms of connections 
an even permutation of un-contracted indices is always
maintained when using the unbarred connection
 so that, for example, an index order
$\not\!\!\alpha\not\!\!\beta\not\!\!\gamma$ can be rearranged as
 $\not\!\!\gamma\not\!\!\alpha\not\!\!\beta$ but not as
$\not\!\!\alpha\not\!\!\gamma\not\!\!\beta$ 
which has an odd number of
indices interchanged. The insertion of a dummy index
does not change this since it still requires an
odd number of interchanges to go from;
$\not\!\!\alpha\not\!\!\beta\not\!\!\phi\not\!\!\gamma\;$ 
to $\;\not\!\!\alpha\not\!\!\gamma\not\!\!\phi\not\!\!\beta$.
We have;
\begin{eqnarray}
&\;&u_{\not{\alpha}\;;\;\not\beta\;;\;\not\gamma}+
u_{\not\alpha\;;\;\not\gamma\;;\;\not\beta}
\\
\nonumber
&=&u_{\not\alpha\;;\;\not\beta\;,\;\not\gamma}
-\Gamma^{\not\phi}_{\not\gamma\not\alpha}u_{\not\phi\;;\;\not\beta}
-\Gamma^{\not\phi}_{\not\beta\not\gamma}u_{\not\alpha\;;\;\not\phi}
+
u_{\not\alpha\;;\;\not\gamma\;,\;\not\beta}
-\Gamma^{\not\phi}_{\not\beta\not\alpha}u_{\not\phi\;;\;\not\gamma}
-\Gamma^{\not\phi}_{\not\gamma\not\beta}u_{\not\alpha\;;\;\not\phi}
\nonumber\\
&=&
u_{\not\alpha\;,\;\not\beta\;,\;\not\gamma}
+\Gamma^{\not\phi}_{\not\alpha\not\beta\;,\not\gamma}u_{\not\phi}
-\Gamma^{\not\phi}_{\not\alpha\not\beta}u_{\not\phi\;,\;\not\gamma}
-\Gamma^{\not\phi}_{\not\gamma\not\alpha}u_{\not\phi\;,\;\not\beta}
+\Gamma^{\not\phi}_{\not\gamma\not\alpha}\Gamma^{\not\rho}
_{\not\phi\not\beta}u_{\not\rho}
-\Gamma^{\not\phi}_{\not\beta\not\gamma}u_{\not\alpha\;,\;\not\phi}
+\Gamma^{\not\phi}_{\not\beta\not\gamma}\Gamma^{\not\rho}
_{\not\alpha\not\phi}u_{\not\rho}
\nonumber\\
&\;&
+
u_{\not\alpha\;,\;\not\gamma\;,\;\not\beta}
+\Gamma^{\not\phi}_{\not\alpha\not\gamma\;,\not\beta}u_{\not\phi}
-\Gamma^{\not\phi}_{\not\alpha\not\gamma}u_{\not\phi\;,\;\not\beta}
-\Gamma^{\not\phi}_{\not\beta\not\alpha}u_{\not\phi\;,\;\not\gamma}
+\Gamma^{\not\phi}_{\not\beta\not\alpha}\Gamma^{\not\rho}
_{\not\phi\not\gamma}u_{\not\rho}
-\Gamma^{\not\phi}_{\not\gamma\not\beta}u_{\not\alpha\;,\;\not\phi}
+\Gamma^{\not\phi}_{\not\gamma\not\beta}\Gamma^{\not\rho}
_{\not\alpha\not\phi}u_{\not\rho}
\nonumber\\
&=&
(\Gamma^{\not\phi}_{\not\alpha\not\beta\;,\not\gamma}
+\Gamma^{\not\phi}_{\not\alpha\not\gamma\;,\not\beta}+
\Gamma^{\not\rho}_{\not\gamma\not\alpha}\Gamma^{\not\phi}
_{\not\rho\not\beta}
+\Gamma^{\not\rho}_{\not\beta\not\alpha}
\Gamma^{\not\phi}_{\not\rho\not\gamma})u_{\not\phi}
\nonumber\\
&=&{\cal{R}}^{\not\epsilon}_{\not\alpha\not\beta\not\gamma}
u_{\not\epsilon}
\label{R}
\end{eqnarray}
where, to obtain the second but last line in eq.(\ref{R})
I have assumed that the vector field is massless and 
employed Proca's equation viz;
\begin{equation}
u_{\not\alpha\;,\;\not\beta\;,\;\not\gamma}+
u_{\not\alpha\;,\;\not\gamma\;,\;\not\beta}
=2I_4\eta_{\beta\gamma}\partial_{\epsilon}
\partial^{\epsilon}u_{\not\alpha}
=0
\label{Procasconstraint}
\end{equation}
and the antisymmetry of the `slashed-gamma'
$\Gamma^{\not\rho}_{\not\beta\not\alpha}=-
\Gamma^{\not\rho}_{\not\alpha\not\beta}$
to define the `curly-R' curvature tensor 
${\cal{R}}^{\not\epsilon}_{\not\alpha\not\beta\not\gamma}$.
This form of the curvature tensor is to be
compared with the conventional form; the latter  is found
by substituting a commutator for an
anticommutator of the covariant derivatives in
eq.(\ref{R}) viz;
\[
u_{{\alpha}\;;\;\beta\;;\;\gamma}-
u_{\alpha\;;\;\gamma\;;\;\beta}=
R^{\epsilon}_{\alpha\beta\gamma}
u_{\epsilon}.
\]
The `curly-R' tensor has many properties similar
to the conventional tensor and those of importance
are proven in the appendix.

Using the Bianci identity and the symmetry of the
contracted `curly-hat' $\hat{\cal{R}}$ 
in its two un-contracted indices 
and  which contains only products of
first 
derivatives of the metric we have the
following;
\begin{eqnarray}
0&=&
g^{\not\beta\not\alpha}
{\hat{\cal{R}}}_{\not\alpha\not\beta\;;\;\not\gamma}
+g^{\not\beta\not\alpha}
{\hat{\cal{R}}}_{\not\beta\not\gamma\;;\;\not\alpha}+
g^{\not\beta\not\alpha}
{\hat{\cal{R}}}_{\not\gamma\not\alpha\;;\;\not\beta}
\nonumber\\
&=&g_s^{\not\beta\not\alpha}
{\hat{\cal{R}}}_{\not\alpha\not\beta\;;\;\not\gamma}+
g^{\not\beta\not\alpha}
{\hat{\cal{R}}}_{\not\beta\not\gamma\;;\;\not\alpha}+
g^{\not\alpha\not\beta}
{\hat{\cal{R}}}_{\not\gamma\not\beta\;;\;\not\alpha}
\nonumber\\
&=&{-1\over2}
{\hat{\cal{R}}}_{\;;\;\not\gamma}+
2g_s^{\not\beta\not\alpha}
{\hat{\cal{R}}}_{\not\beta\not\gamma\;;\;\not\alpha}
\nonumber\\
&=&{-1\over2}
{\hat{\cal{R}}}_{\;;\;\not\gamma}-
{\hat{\cal{R}}}^{\not\alpha}_{\;\;\;\not\gamma\;;\;\not\alpha}
\end{eqnarray}
which allows us to write an `Einstein-like' equation;
\begin{equation}
({1\over2}
g^{\not\alpha\not\gamma}
{\hat{\cal{R}}}
+{\hat{\cal{R}}}^{\not\alpha\not\gamma}
)_{\;;\;\not\gamma}=0
\label{Eino}
\end{equation}
Unlike the actual Einstein equation however
the bracketed part of eq.(\ref{Eino})
is traceless as follows. Lowering the
$\not\!\!\alpha$ index and tracing;
\[
{1\over2}
g_{\not\gamma}^{\;\;\;\not\gamma}
{\hat{\cal{R}}}
+{\hat{\cal{R}}}_{\not\gamma}^{\;\;\;\not\gamma}
={-\hat{\cal{R}}}+{\hat{\cal{R}}}=0
\]
because $g_{\not\gamma}^{\;\;\not\gamma}
=
-{I_4\over2}\eta_{\not\gamma}^{\;\;\not\gamma}=-2I_4$.
We wish to equate eq.(\ref{Eino})
to the vanishing divergence of a 
stress-energy tensor $T^{\not\alpha\not\gamma}
_{\;\;\;\;\;;\not\gamma}=0$. It is clear that this
equation must be satisfied independently for
any symmetric and anti-symmetric components of 
the stress-energy tensor (we may reasonably
assume the anti-symmetric part of the stress-energy
tensor is zero) and thus equating symmetric parts
immediately leads to; 
\[
(+{1\over2}
g_s^{\not\alpha\not\gamma}
{\hat{\cal{R}}}
+{\hat{\cal{R}}}^{\not\alpha\not\gamma}
)_{\;;\;\not\gamma}=0={T_s^{\not\alpha\not\gamma}}
_{\;;\;\not\gamma}\]
Replacing the metric $g_s^{\not\alpha\not\gamma}$
by its constant  symmetric part 
${-I_4\over2}\eta^{\alpha\gamma}$,
and noting that because ${\cal{R}}^{\not\alpha\not\gamma}$
is symmetric  we may drop the slash notation 
on the symmetric tensors and multiply through by the constant
${1\over2}\gamma^{\gamma}$ 
to eliminate the slash on the derivative
 index to obtain;
\begin{equation}
I_4(-{1\over4}
\eta^{\alpha\gamma}
{\hat{\cal{R}}}
+{\hat{\cal{R}}}^{\alpha\gamma}
)_{\;;\;\gamma}=0=c
{I_4T^{\alpha\gamma}}_{\;;\;\gamma}
\label{Eino2}
\end{equation}
which we recognize, dropping the $I_4$'s,
 as the correct form for the 
stress-energy equation for a free massless field.
(Of course the L.H.S. of eq.(\ref{Eino2}) -
the curvature with un-slashed co-ordinates 
- is only meaningful if
referred to a `spinorial' observer
observing in anti-commuting co-ordinates).
The constant c must have dimension $l^{2}$
because the stress-energy tensor has the
dimension $l^{-4}$ and the curvature tensor
has the dimension $l^{-2}$.
Thus the scalar curvature, and Lagrangian, is given by;
\begin{equation}
k\,{\hat{\cal{R}}}
=-{1\over4}F_{\mu\nu}
F^{\mu\nu}=+{1\over4}F_{\mu\nu}F^{\nu\mu}
\end{equation} 
where the  constant k is of dimension 
of inverse  length
squared. This constant will not be 
related to the gravitational constant
as such but must be a function of the
length scale of the source 
over which the charge is distributed.
As a reasonable ansatz we may take;
\begin{equation}
k\propto{\sqrt{|g|}}
\;\alpha^2m_0^2
\end{equation}
where $\alpha$ is the fine-structure
constant,
$|g|$ is the modulus of the determinant
of the metric 
 and $m_0$ is the rest mass of
the source.

\section{Potential and Source terms}

If the speculation entered into at the beginning of
this paper is valid we expect to find curved anti-commuting
co-ordinates manifestly in the vicinity of charged
particles such as an electron or proton. Imagine,
with at least some degree of creative license,
that you are an observer sitting on an electron.
What does the space-time in your vicinity look
like? The  thing of note is that, if you
rotate by $2\pi$ around a fixed axis, your 
co-ordinates anti-commute. Your space-time is
not the same as expected for a macroscopic
`bosonic' observer. Thus we actually expect
the anti-commuting space-time to dominate in
the vicinity of a fermion; indeed, the fermion 
itself must be actually generating this aspect of
local space-time - it must be the source. (Analogously
we must expect the gravitating masses of the
universe to be, not just the source of the
gravitational field, but the source of
commuting space-time itself if this scenario is
valid).

We want to interpret this anti-commuting aspect of space-time,
superimposed upon gravity-generated commuting space-time,
as the electric field of the charged particle. 
We can encompass two kinds of charges with
anti-commuting space-time because, unlike with
the case of the usual commuting co-ordinates,
  complex-conjugate  types of
space-time in anti-commuting co-ordinates
can be defined;
$x^{\not\alpha}$ and $x^{\dagger\not\alpha}
\equiv
{1\over2}x^{\alpha}\gamma^{\dagger\alpha}$
(no sum on $\alpha$).
The theory is symmetric between these two
so that, for example, we have the hermitian
conjugate metric coupling to the hermitian
conjugate co-ordinates;
\[
\bar{g}^{\dagger}_{\not\alpha\not\beta}
x^{\dagger\not\beta}
=x_{\not\beta}
\bar{g}^{\not\beta\not\alpha}
=x^{\not\alpha}
=
x_{\not\alpha}^{\dagger}
\]
and so-on for the other relations given in the
appendix. Since $x^{\alpha}{\neq}x^{\dagger\alpha}$
there are in fact two possible types of anti-commuting 
space-times  which we will associate with the two
possible charge types that exist in the universe;
positive and negative. But note that, in the given
representation, they do not differ in their
time representation since $\gamma^0=\gamma^{\dagger0}$.
 
The question naturally arises as to which type of
space-time, commuting or anti-commuting,
 dominates in the quantum domain. Dominates is 
perhaps not quite the correct term
here as the issue is the `stiffness' of the
space-time. Because of the `stretching' of the 
commuting space-time with respect to
anti-commuting space-time with  expansion of the universe
we expect the coupling constants of the two forces
to be related as a function of the degree of
`stretching' of space-time. Thus commuting space-time
is more than 30 orders of magnitude `stiffer' than
anti-commuting space-time.
 Thus in the quantum domain 
in the current epoch we expect the
anti-commuting space-time to be  much more
`flexible' than the commuting space-time.
 Only in the very early history of the
universe when the coupling constants of gravitation
and electro-magnetism are more equal
 does the commuting space-time begin
to be in parity with the anti-commuting space-time.
But note that the cancelation between different
charged space-times only occurs for the space dimensions;
\begin{equation}
\gamma^{i}+\gamma^{{\dagger}i}=
\gamma^{i}-\gamma^{i}=0\;\;;\;\;
\gamma^{0}+\gamma^{{\dagger}0}=
2\gamma^{0}
\end{equation}
and the anti-commuting time-part of space-time
will expand with the commuting space-time 
and the observed macroscopic time dimension is then
the sum of two parts; one generated by the commuting
part of space-time and one generated by the anti-commuting 
part. 

Of course this also means that the time dimension
in anti-commuting space-time in the vicinity of a 
charged particle is `stiff'; because it is stretched
like the commuting space-time it does not bend easily
- only the space dimensions are significantly bent
in the region of a charged particle and the time is
effectively `flat'. Thus the symmetry of the charged object
with respect to its electro-magnetic field  may
be represented, to a good approximation, by a
compact space. Is it  perhaps for this reason
that spinorial particles are represented
by a compact representation of the L\"{o}rentz group;
SU(2) is in fact a representation of SO(3)
up to a sign? Note that the stiffness of the time
dimension does not effect the description of
free-fields since the photon has no longitudinal
or time-like polarization (any electro-magnetic
bending of the `stiff' time dimension is suppressed
by more than 30 orders of magnitude).

Putting aside the issue of the dual nature
of the time dimension in this theory
we may ask how might it be possible to
incorporate the basic features of quantum
physics into the metric structure to 
describe the source? 
The non-vanishing $x^{\not\,0}$ component of the
vacuum remote from charged sources would not appear to
effect the free-field tensor since the longitudinal
polarization of a photon is zero and the
photon does not cause oscillations in the `stiff'
time dimension (or at least these are suppressed
to the order of the gravitational coupling constant).
 However, we may
modify the metric to represent the anti-commuting
space-time in the vicinity of the source 
by neglecting the time co-ordinate and
treating the space as compact. Consider the 
following metric;
\begin{equation}
g_{{\not}\,i{\not}\,j}={1\over2}e^{i{{2\pi}\over\hbar}p{\cdot}x}
(\delta_{{\not}\,i{\not}\,j}+i\sigma_{{\not}\,i{\not}\,j})
=\bar{g}_{{\not}\,j{\not}\,i}
\label{uncertainty}
\end{equation}
where $p{\cdot}x=p_{\not\epsilon}x^{\not\epsilon}
=p_{\epsilon}x^{\epsilon}$ and $p_{\epsilon}$
is the particle four-momentum.
(The index raising form is made
by setting p=0).  Now the wave factor introduced
into the metric means that the distance 
$ds^2=Tr.dx^{\not\alpha}\bar{g}
_{\not\alpha\not\beta}dx^{\not\beta}$
  oscillates in 
value between positive and negative values in
this case of {\it{spatial distance}} since,
although the wave factor contains a $p_0x^0$
 term, it is part of a scalar which 
may be referred to the commuting
co-ordinates as above. We assume that such a metric
is valid in the immediate vicinity  of a source.
Because the metric is complex and referred
to complex co-ordinates $ds^2$  never vanishes
in spite of the phase-factor. 
By the principle of relativity 
this should also be true for the observer with
respect to
commuting co-ordinates. If the imaginary part of the
 metric is unmeasurable (that is,
if the real part only is measurable) then the
imaginary part may be represented by the uncertainty
of the measurement. Now the phase factor oscillates
with both real and imaginary components and the
metric (\ref{uncertainty}) contains a real part
$\eta_{i\,j}$ and a pure imaginary part $i\sigma_{i\,j}$.
Now;
\[|i\sigma_{i\,j}|^{\,2}=6\;\mbox{and}\;|\eta_{i\,j}|^{\,2}=3\]
so that the imaginary part of metric (\ref{uncertainty})
is maximised when the phase factor is purely real.
Thus the imaginary part of the metric is maximised
for the corresponding angle in the phase
factor; \begin{equation}
{2{\pi}n\over\hbar}p{\cdot}x=2{\pi}n
\label{lid}\end{equation}
where n is a constant  integer (we exclude the trivial case
where $n=0$). The measurement of the real part of the
corresponding product will thus have an equal or
greater uncertainty in its  measurement than the 
value specified by  eq.(\ref{lid}). We thus
obtain;

\begin{equation}
|
{\bigtriangleup}p{\cdot}{\bigtriangleup}x
|\geq{\hbar}
\end{equation}

Thus  the presence of a source phase-factor representing
a wave-function in the metric related
to the anti-commuting part of space-time 
may be interpreted as an expression of
the uncertainty principle when referred to commuting
co-ordinates (i.e. the observers co-ordinate system).
 That the uncertainty only applies
in the quantum scale of things results from the
effective vanishing of anti-commuting space-time
over macroscopic distances.

Armed with this idea can we extract an expression
for a charged source tensor from metric (\ref{uncertainty})?
For starters we note that the given derivation
of the `curly-$\hat{\cal{R}}$'
 curvature tensor breaks-down; it is predicated
on an object  satisfying a massless Proca's equation and
which also satisfies the L\"{o}rentz condition (although
the latter is actually imposed by the mathematical
constraints of the theory; see appendix). That is,
the derivation of the free-field curvature
tensor has relied on a massless vector field
whilst the sources are massive spinors. 
It is also predicated on the vanishing of the
derivative of the symmetric part of the
metric; and this condition breaks down with
the insertion of a phase factor
in metric (\ref{uncertainty}).
To study spinors
the mathematical machinery must be re-geared
and this  has not yet been successfully
achieved. 

We might nevertheless attempt an
ansatz based on general ideas.
The key difference between the
`Einstein-like' conservation equation
resulting from the contraction of the
Bianci identity for the `curly-R' 
curvature tensor and that for General
Relativity is the tracelessness of
the former. In the presence of a source
the trace of the equation will not
vanish and we require an equation of
the form (for convenience dropping
the slash notation under the assumption
that the equation is understood to
imply curvature of the anti-commuting
part of space-time);

\begin{equation}
k(-{1\over4}\eta^{\alpha\beta}\hat{\cal{R}}_{FF}
+\hat{\cal{R}}^{\alpha\beta}_{FF})+
k^{'}\hat{\cal{R}}_{S}^{\alpha\beta}
=T^{\alpha\beta}
\end{equation}

\noindent
where $\hat{\cal{R}}_S^{\alpha\beta}$
is a curly-R type curvature tensor for
anti-commuting co-ordinates and
contains analogous components from the
contraction of a Bianci identity
related to a source metric with non-vanishing
derivatives (it may contain a scalar piece
multiplied by the metric for example).
The new constant $k\,'$ here is, of course,
of dimension $l^{-2}$.

Of course to construct ${\hat{\cal{R}}}_S$
we must replace the massless vector field
in eq.({\ref{Procasconstraint}})    by a massive spinor field in
the derivation of the curvature tensor;

\begin{equation}
u_{\not\alpha\;,\;\not\beta\;,\;\not\gamma}+
u_{\not\alpha\;,\;\not\gamma\;,\;\not\beta}
=2I_4\eta_{\beta\gamma}\partial_{\epsilon}
\partial^{\epsilon}u_{\not\alpha}
=2g^s_{\not\beta\not\gamma}m^2_0u_{\not\alpha}
\label{Procasconstraint2}
\end{equation}

\noindent
so that a constant multiple of the mass squared
is added to the definition of the curvature
tensor derived from the anti-commutator of
eq.(\ref{R}). One might guess from this
that the appropriate addition 
to the stress-energy tensor 
for the free-field would then be
a constant multiple of the mass-squared
and for the vector piece the product
of momenta $p_{\not\alpha}p_{\not\beta}$
which, with dropping of the slash
notation and appropriate adjustment 
of constants, can be converted into the
conventional source stress-energy tensor
for a charged particle {\cite{CFT}.

 Because of the
symmetry of the contracted curly-R curvature
tensor in its two un-contracted indices for this to 
be a viable approach we will require that 
the momenta commute so that this
can be re-written as $2a.p_{\not\beta}p_{\not\gamma}$;
however one expects the momenta in anti-commuting
co-ordinates to anti-commute so that this 
tensor reduces to a trivial result.
An alternative may be to consider the analogue
of conventional curvature  in anti-commuting
space-time to feed
into a Bianci identity to construct a conserved
quantity;

\begin{equation}
\check{\cal{R}}^{\;\;\not\alpha}_{s\not\gamma\not\beta\not\alpha}=
\Gamma^{\not\epsilon}_{\not\beta\not\gamma}\,\Gamma^
{\not\alpha}_{\not\epsilon\not\alpha}+
\Gamma^{\not\epsilon}_{\not\alpha\not\gamma}\,\Gamma^
{\not\alpha}_{\not\epsilon\not\beta}
\end{equation}

\noindent
where once again I have dropped the terms involving
second derivatives as they will vanish. Note that 
for metric eq.(\ref{uncertainty})
the derivative of the symmetric part of the
metric does not vanish and the connection is
not completely anti-symmetric. The tensor
$\check{\cal{R}}^{\;\;\not\alpha}
_{s\not\gamma\not\beta\not\alpha}$
will thus be expected to  contain both a symmetric
and an anti-symmetric part or more generally 
be expressible as the sum of two tensors;
one symmetric and one anti-symmetric. We may 
expect, with regard to the source
momenta, that  the symmetric component will be 
trivial and so the conserved quantity
will be then expressed as an anti-symmetric
tensor which must be coupled to a
stress-energy tensor which is also anti-symmetric.
Conventional wisdom dictates that such a 
stress-energy tensor will not conserve
angular momentum but this may not be the
case here for spinors in anti-commuting
space-time. The resolution of these issues,
if indeed they can be resolved, must await
development of a means of adapting the
derivation of curly-R curvature tensors from
spinor fields; which as mentioned is work
still outstanding.

\section{Conclusion}

The reader at this stage may well object,
with  some justification, ``what has been achieved
by this construction since we have a good
theory of electro-magnetic processes in the
form of Q.E.D. to which the above theory adds
no new prediction and cannot, at least at its
current level of development,
 match current theory?''
It is difficult to disagree with this complaint
except to add that  at some point one must try to incorporate
gravity into the scheme of things since
its omission is clearly a physical impossibility.
This new approach however does hold some promise
of new physical predictions at extreme energies
where the coupling constants of gravitation and
electro-magnetism are comparable. Also the
ability to give an account of the wide
difference in the coupling constants of the
two forces at low energy is a useful feature
and somthing unaccounted for in standard
theory (Kaluza-Klein theory not withstanding).
The ability to give an account for the
origin of the uncertainty principle on the
basis of pure geometry seems to be a new
feature in physics as far as I can ascertain.

Attempts at geometric unification are of course nothing new
dating back to the work of N{\"{o}}rdstrom, Weyl,
Kaluza-Klein, Eddington, Cartan and Einstein.
(For a review of the history of the subject
see \cite{russian}; for more recent work in a similar
vein to the Weyl-Eddington approach see \cite{WE}.
Standard works involving asymmetric
tensor fields can be found in  Einstein \cite{UFT},
Kibble and Sciami \cite{T},\cite{3},\cite{4}.
For more recent extensions to Yang-Mills fields see
\cite{Mielke}, \cite{Lunev}, \cite{Haag} and \cite{Rad}
and for an extensive  recent review of the
subject see \cite{F.W.}.
Note however that  something rather different is 
being developed in this current paper
to most previous attempts; the coupling
of the anti-symmetric connection to the
anti-symmetric co-ordinate system means that 
the connection does not represent torsion).

This work should then be seen in this light;
Q.E.D. clearly contains profound truth - its
empirical validity tells us this - but neither
it nor its electro-weak (standard model) extension
can be complete. To couple it to gravitation
we need something completely new. The present
work follows on from many previous attempts
but avoids the problems associated with
introduction of a torsion field; for
anti-commuting co-ordinates the anti-symmetric
connection propagates;
\begin{equation}
\Gamma^{\not\alpha}_{[\not\beta\,\not\gamma{]}}
{dx^{\not\beta}\over{ds}}{dx^{\not\gamma}\over{ds}}
\neq
0
\end{equation}
unlike the case for commuting co-ordinates.
(In which case it is interpreted as a spin-contact
interaction - see \cite{four} and \cite{Hehl}).
Mixing of commuting and anti-commuting
space-time results in a new symmetry.
This new symmetry is encapsulated in the
expression of both gravitation and
electro-magnetism in terms of curvature
of space-time; in the case of gravitation
the curvature is of the commuting part of space-time
and in the case of electro-magnetism of the
anti-commuting part.  The expression of
both gravitation and electro-magnetism
in terms of general co-ordinate
transformations implies
invariance of the laws of nature with respect to 
arbitrary mixtures of commuting and anti-commuting
co-ordinates although no formal proof of this statement
has yet been developed; the laws of the universe 
should be valid for general co-ordinate transformations
in the current epoch and at all preceding and
succeeding times as the mix will vary with the
age of the universe if the suppositions presented in
this paper are correct.

On the other hand one may ask what happens
to particle statistics
when, as in the early universe, commuting space-time
is as flexible as anti-commuting space-time?
Is the electron a spinor precisely because its
major physical manifestation is the generation
and bending of adjacent anti-commuting space-time?
And if this is the case shouldn't it manifest bosonic
characteristics if the commuting space-time it is
generating reaches equal flexibility with the 
anti-commuting space-time near the Planck scale?

It is possible that these conclusions are a
consequence of the theory; that is, that 
objects such as electrons manifest continuous statistics
which only becomes apparent at very high energy. 
However it is also possible that the continuous statistics
require quantization; by which is meant that, instead
of continuous statistics for an object like an electron,
one encounters a new particle at high energy which
is a bosonic super-partner of the electron that  `carries
off' the `bosonic' part of the electron statistics
at high energy. In that case
there would be a more conventional type of supersymmetry
(if indeed supersymmetry may be called conventional).

A comment on  the issue of the principle
of equivalence is appropriate as it 
 is so much at the center
of General Relativity. It is represented
by the vanishing of the covariant derivative
of the metric and this, in turn, tells us that
a frame exists in anti-commuting space-time
in which  the electro-magnetic
field  can be `transformed away'
{\it{locally}}; the analogue of a free-fall frame
in commuting space-time. This in turn means that the
physics can be described by curvature in the anti-commuting
co-ordinates.

But what are we to make of the idea of the equivalence
of charge and mass expected near the Planck scale? 
 We
know that charge comes in fixed quantum
integral multiples of $1\over3$ but particle
masses vary considerably for the same
unit of charge. Both mass and charge
vary with energy but not in the same
way; the coupling constant of the
E.M. field $\alpha\approx137^{-1}$
varies with energy but not in a way that
is easily related to a L\"{o}rentz boost by
a multiplicative constant (it takes 
the value $\approx128^{-1}$ at the weak
unification scale and is thought to take
a value order unity  at the Planck scale).
Clearly the symmetry between mass and 
charge is badly broken in nature. If the
central surmise of this paper is correct
the symmetry would be exact only in the
situation that the universe consisted
solely of a single  electron with  photons,
and  gravitons  ($\pm$ their corresponding 
supersymmetry partners?). The universe is
however much more complex than this. The
existence of the weak interaction means that
the concept of charge is related to a more
complex vacuum structure than that represented
in this paper leading to massive electro-magnetically
charged non-composite 
fundamental bosons. (The vacuum of the
presented theory is represented
by the effective vanishing of the $x^{\not{\,i}}$
anti-commuting
co-ordinates in regions remote from charged
sources; if these are non-vanishing the principle of
equivalence of charge and mass means that
the vacuum energy is altered - it also
means that the intrinsic statistics of the
vacuum are altered. One intriguing possibility
is that the rate of expansion of the $x^{\not{\,0}}$
time component is slightly retarded with
respect to $x^0$ embedding a direction in the time
dimension and modifying the vacuum structure). Moreover
the Higgs field alters the masses of the particles
as does the strong interaction - which introduces
a further complication in terms of color charge. However, the 
most important physical feature which appears
to break the symmetry between charge and mass
is the stretching of the commuting space-time
due to expansion of the universe and more 
specifically the observation that
the anti-commuting time dimension stretches
like the commuting time dimension leading to a
profound disturbance of any
 simple symmetry between commuting and anti-commuting
space-time and hence between mass and charge. The
charge ends up related to a compact group and the
mass to a non-compact one.

Lastly the deficiencies of the current approach
need to be admitted. The mathematical structure is
predicated on a massless spin-one field. Much work 
remains to extend the theory to massive spinors for successful
incorporation of sources. Nevertheless there is
enough scope in the theory to hope that a way
may be found in the future. The theory may also have
potential for extension to non-abelian Yang-Mills fields
but at the moment this is lacking.

\appendix
\section{Notation and Conventions}
The co-ordinates are coupled to
4x4 gamma matrices and represented with a `slash'
notation;
\[
x_{{\not}\,0}\equiv
x_0{\gamma_0\over2}
\;,\;
x_{{\not}\,1}\equiv
x_1{\gamma_1\over2}
\;,\;
x_{{\not}\,2}\equiv
x_2{\gamma_2\over2}
\;,\;
x_{{\not}\,3}\equiv
x_3{\gamma_3\over2}
\]
Where the 4x4 gamma matrices satisfy;
\[ [\gamma_{\alpha},\gamma_{\beta}]_-
=
2I_4\eta_{\alpha\beta}
\;
\mbox{and}
\;
\eta_{\alpha\beta}=\mbox{diag.}(+,-,-,-)_{\alpha\beta}
\]
Indices are raised or lowered by the 16x16 metric;
\[
g_{\not\alpha\not\beta}
=
-{1\over2}(I_4\eta_{\alpha\beta}-i\sigma_{\alpha\beta})
;\;
,
\;\;
\bar{g}_{\not\alpha\not\beta}
=
-{1\over2}(I_4\eta_{\alpha\beta}+i\sigma_{\alpha\beta})
\]
and hence $\bar{g}_{\not\alpha\not\beta}
={g}_{\not\beta\not\alpha}$,
and we have;
\begin{equation}
\bar{g}_{\not\beta\not\alpha}x^{\not\alpha}=x_{\not\beta}
\;,\;
{g}_{\not\beta\not\alpha}x^{\not\alpha}=-2x_{\not\beta}
\;,\;
{g}_{\not\beta\not\alpha}x^{\not\alpha^{\dagger}}=x_{\not\beta}
\;,\;
\bar{g}_{\not\beta\not\alpha}x^{\not\alpha^{\dagger}}
=-2x_{\not\beta}
.
\label{prop}
\end{equation}
Metric products have the following properties;
\[
g_{\not\alpha\not\beta}g^{\not\beta\not\epsilon}
=
\delta_{\not\alpha}^{\;\;\not\epsilon}
=
\bar{g}_{\not\alpha\not\beta}\bar{g}^{\not\beta\not\epsilon}
=
I_4\delta_{\alpha}^{\;\;\epsilon}
\]
\[
\bar{g}_{\not\alpha\not\beta}{g}^{\not\beta\not\epsilon}
=
g_{\not\alpha}^{\;\;\not\epsilon}\;\;;\;\;
{g}_{\not\alpha\not\beta}{g}^{\not\epsilon\,\not\beta}
=
g_{\not\alpha}^{\;\;\not\epsilon}
\]
\section{TRANSFORMATION PROPERTIES OF THE METRIC}

Gamma matrices and sigma matrices transform under 
L\"{o}rentz transformations as follows;

\[
S^{-1}\gamma^{\mu}S=\Lambda^{\mu}_{\;\;\nu}\gamma^{\nu}
\;\;\;\;\mbox{and}\;\;\;\;
S^{-1}\sigma^{\mu\nu}S=\Lambda^{\mu}_{\;\;\lambda}
\Lambda^{\nu}_{\;\;\kappa}\sigma^{\lambda\kappa}
\]
where $S{\approx}(1+{i\over2}\epsilon^{\mu\nu}
\sigma_{\mu\nu})$
and
$\Lambda^{\nu}_{\;\;\nu}{\approx}(g^{\mu}_{\;\;\nu}+
\epsilon^{\mu}_{\;\;\nu})
$
with the infinitesimal $\epsilon^{\mu\nu}$ 
 antisymmetric in 
$\mu$ and $\nu$. Neither the gamma matrices nor the 
sigma matrices commute with the spinor representation of
the L\"{o}rentz group S so that neither transforms as an
invariant  tensor 
under the L\"{o}rentz group. However, the modification of the
metric by the inclusion of an antisymmetric $\sigma$ matrix term 
{\it{when coupled to anti-commuting co-ordinates}}
does transform appropriately as we will now see. The
generalization to gamma-slashed co-ordinates is (to
first order in infinitesimal);
\begin{eqnarray}
\Lambda^{\mu}_{\;\;\nu}\rightarrow
{\Lambda}^{\not\mu}_{\;\;\not\nu}\approx
(g^{\not\mu}_{\;\;\not\nu}+\epsilon^{\not\mu}_{\;\;\not\nu})
\label{A1}
\end{eqnarray}
where g now includes both a symmetric and an antisymmetric
part. A bar  $\bar{\Lambda}$ is as per the metric
bar operation and
indicates hermitian conjugation plus
 raising or lowering of gamma-matrix coupled indices so that;
\[
\overline{\epsilon_{\not\alpha}^{\;\;\;\not\beta}}
=
({\epsilon^{\not\alpha}_{\;\;\;\not\beta}})^{\dagger}
=
{\epsilon_{\;\;\;\not\alpha}^{\not\beta}}
=
-{\epsilon_{\not\alpha}^{\;\;\;\not\beta}}
.\]
 The co-ordinates transform
as (from properties \ref{prop});
\[
x^{\not\alpha}\Lambda^{\not\alpha^{'}}
_{\;\;\;\not\alpha}=x^{\not\alpha}
\bar{\Lambda}^{\;\;\;\not\alpha^{'}}
_{\not\alpha}=x^{\not\alpha^{'}}
\;\;\mbox{and}\;\;
\bar{\Lambda}^{\not\alpha^{'}}_{\;\;\;\not\alpha}
x^{\not\alpha}
={\Lambda}^{\;\;\;\not\alpha^{'}}_{\not\alpha}
x^{\not\alpha}
=x^{\not\alpha^{'}}
\]
Now
defined as
(to first order in the infinitesimals);
\begin{eqnarray}
\bar{\Lambda}_{\not\alpha^{'}}^{\;\;\not\alpha}
{g}_{\not\alpha\not\beta}
{\Lambda}_{\not\beta^{'}}^{\;\;\not\beta}
&=&
(\bar{g}_{\not\alpha^{'}}^{\;\;\not\alpha}
+\epsilon_{\not\alpha^{'}}^{\;\;\not\alpha})
g_{\not\alpha\not\beta}
({g}_{\not\beta^{'}}^{\;\;\not\beta}
+\epsilon_{\not\beta^{'}}^{\;\;\not\beta})
\\
\nonumber
&=&
(\bar{g}_{\not\alpha^{'}}^{\;\;\not\alpha}
+\epsilon_{\not\alpha^{'}}^{\;\;\not\alpha})
(g_{\not\alpha\not\beta^{'}}
+2\epsilon_{\not\alpha\not\beta^{'}})
\\
\nonumber
&=&
g_{\not\alpha^{'}\not\beta^{'}}
-2\epsilon_{\not\alpha\not\beta^{'}}
+2\epsilon_{\not\alpha\not\beta^{'}}
\\
\nonumber
&=&
g_{\not\alpha^{'}\not\beta^{'}}
\end{eqnarray}
and similarly one can show that
$\bar{\Lambda}_{\not\alpha^{'}}^{\;\;\not\alpha}
\bar{g}_{\not\alpha\not\beta}
{\Lambda}_{\not\beta^{'}}^{\;\;\not\beta}
=
\bar{g}_{\not\alpha^{'}\not\beta^{'}}$
so that we obtain the relations;
\begin{eqnarray}
ds^2&=&Tr.dx^{\not\alpha^{'}}dx_{\not\alpha^{'}}
=Tr.dx^{\not\alpha}
\Lambda^{\not\alpha^{'}}
_{\;\;\;\not\alpha}
\bar{g}_{\not\alpha^{'}\not\beta^{'}}
\bar{\Lambda}^{\not\beta^{'}}
_{\;\;\;\not\beta}
dx^{\not\beta}
=
Tr.dx^{\not\alpha}\bar{g}_{\not\alpha\not\beta}
dx^{\not\beta}=ds^2
\nonumber\\
ds^2&=&
(Tr.g)^{-1}Tr.dx^{\not\alpha}
\bar{\Lambda}^{\;\;\;\not\alpha^{'}}
_{\not\alpha}
{g}_{\not\alpha^{'}\not\beta^{'}}
{\Lambda}^{\;\;\;\not\beta^{'}}
_{\not\beta}
dx^{\not\beta}
=
(Tr.g)^{-1}Tr.dx^{\not\alpha}{g}_{\not\alpha\not\beta}
dx^{\not\beta}=ds^2
\end{eqnarray}
(Note; the inverse trace g in the last expression 
occurs because of metric contraction properties
\ref{prop} and has the constant value $-{1\over2}$).
The metric thus acts as an invariant
 tensor under the {\it{modified}}
co-ordinate system and $ds^2$ is an invariant.

\section{Proof of the L\"{o}rentz identity}
This applies to the
free-field in which case we
set the derivative of the 
symmetric part of the metric to zero
i.e. $\eta_{\not\alpha\not\beta\;,\;\not\gamma}
=0\;\;;\;\{\forall
\;\not\gamma\}$.
Using the identities;
\[
g_{\not\alpha\not\phi}\bar{g}^{\not\phi}_{\;\;\not\beta}
=g_{\not\alpha\not\beta}\;\;
\mbox{and}
\;\;
\bar{g}_{\not\alpha}^{\;\;\not\phi}g_{\not\phi\not\beta}
=
\bar{g}_{\not\alpha\not\beta}=g_{\not\beta\not\alpha}
\]
we have;
$
g_{\not\alpha\not\beta\,,}^{\;\;\;\;\;\;\;\not\alpha}
=
g_{\not\alpha\not\beta\;,}^{\;\;\;\;\;\;\;\not\phi}
\;\bar{g}_{\not\phi}^{\;\;\not\alpha}
=-{i\over2}\sigma_{\alpha\beta\;,}^{
\;\;\;\;\;\;\;\not\phi}\;
\bar{g}_{\not\phi}^{\;\;\not\alpha}\;\;$
and also$\;\;
g_{\not\alpha\not\beta\,,}^{\;\;\;\;\;\;\;\not\alpha}
=g_{\not\alpha}^{\;\;\;\not\phi}
\bar{g}_{\not\phi\not\beta\;,}^{\;\;\;\;\;\;\;\not\alpha}
=
\bar{g}^{\not\alpha}_{\;\;\;\not\phi}
g_{\not\beta\not\alpha\;,}^{\;\;\;\;\;\;\;\not\phi}
=+{i\over2}
\bar{g}^{\not\alpha}_{\;\;\;\not\phi}
\sigma_{\alpha\beta\;,}^{
\;\;\;\;\;\;\;\not\phi}\;
$whence;

\begin{eqnarray}
-{i\over2}\sigma_{\alpha\beta\;,}^{
\;\;\;\;\;\;\;\not\phi}\;
\bar{g}_{\not\phi}^{\;\;\not\alpha}\;\;
-
{i\over2}
\bar{g}^{\not\alpha}_{\;\;\;\not\phi}
\sigma_{\alpha\beta\;,}^{
\;\;\;\;\;\;\;\not\phi}\;
=&0&
\nonumber\\
\;\mbox{or;}\;\;\;\;\;
\sigma_{\alpha\beta\;,}^{
\;\;\;\;\;\;\;\not\phi}\;
(
\bar{g}_{\not\phi}^{\;\;\not\alpha}
+g_{\not\phi}^{\;\;\not\alpha})
=
\sigma_{\alpha\beta\;,}^{
\;\;\;\;\;\;\;\not\phi}\;
I_4\eta_{\phi}^{\;\;\;\alpha}
=
\sigma_{\alpha\beta\;,}^{
\;\;\;\;\;\;\;\not\alpha}\
=&0&
\end{eqnarray}
where the second line follows since,
with vanishing of the derivative of
any symmetric component generated
in the commutation of the sigma matrices
and the picking up of an extra minus sign
from the commutation of the sigma matrix
with the derivative index $\not\!\partial
^{\phi}$ we get;
\begin{eqnarray}
\sigma_{\alpha\beta\;,}^{\;\;\;\;\;\;\;\not\phi}
\bar{g}_{\not\phi}^{\;\;\;\not\alpha}
&=&
\sigma_{\alpha\beta\;,}^{\;\;\;\;\;\;\;\not\phi}
{1\over2}
I_4(\eta_{\phi}^{\;\;\;\alpha}
-i\sigma_{\phi}^{\;\;\;\alpha})
\\
\nonumber
&=&
{1\over2}
(I_4\eta_{\phi}^{\;\;\;\alpha}
\sigma_{\alpha\beta\;,}^{\;\;\;\;\;\;\;\not\phi}
-i\sigma_{\alpha\beta\;,}^{\;\;\;\;\;\;\;\not\phi}
\sigma_{\phi}^{\;\;\;\alpha})
\\
\nonumber
&=&
{1\over2}
(I_4\eta_{\phi}^{\;\;\;\alpha}
\sigma_{\alpha\beta\;,}^{\;\;\;\;\;\;\;\not\phi}
-i\sigma_{\phi}^{\;\;\;\alpha}
\sigma_{\alpha\beta\;,}^{\;\;\;\;\;\;\;\not\phi})
\\
\nonumber
&=&\bar{g}_{\not\phi}^{\;\;\;\not\alpha}
\sigma_{\alpha\beta\;,}^{\;\;\;\;\;\;\;\not\phi}
\end{eqnarray}
Where the slash notation has been dropped
when the metric is expanded out into
its constituent symmetric and anti-symmetric
parts.
The L\"{o}rentz condition ensures that for 
any derivative of the free field component
e.g. $g^A_{\not\alpha\not\beta\;,\;\not\gamma}$ the 
three indices must label different values.
 i.e. $\alpha\neq\beta\neq\gamma\neq\alpha$. 

\section{Curvature tensor identities}
In the case that the derivatives of the symmetric
part of the metric vanish the connection is 
anti-symmetric in its lower two indices (a torsion
connection in commuting
co-ordinates but not in anti-commuting 
co-ordinates). The antisymmetry of the connection
leads to the following identities.

The contracted curvature tensor is
symmetric in its un-contracted indices;
\begin{equation}
{\cal{R}}^{\not\alpha}_{\not\alpha\not\beta\not\gamma}
={\cal{R}}_{\not\beta\not\gamma}
={\cal{R}}_{\not\gamma\not\beta}
\end{equation}
which follows directly from eq.(\ref{R}).

The curvature tensor (and its barred analogue
formed from the barred form of the connection) 
has the following
cyclic property;
\begin{equation}
{\bar{\cal{R}}}^{\not\epsilon}_{\not\alpha\not\beta\not\gamma}
+
{\bar{\cal{R}}}^{\not\epsilon}_{\not\beta\not\gamma\not\alpha}
+
{\bar{\cal{R}}}^{\not\epsilon}_{\not\gamma\not\alpha\not\beta}=
{\cal{R}}^{\not\epsilon}_{\not\alpha\not\beta\not\gamma}
+
{\cal{R}}^{\not\epsilon}_{\not\beta\not\gamma\not\alpha}
+
{\cal{R}}^{\not\epsilon}_{\not\gamma\not\alpha\not\beta}=0
\end{equation}
viz;
\begin{eqnarray}
{\cal{R}}^{\not\epsilon}_{\not\alpha\not\beta\not\gamma}
+
{\cal{R}}^{\not\epsilon}_{\not\beta\not\gamma\not\alpha}
+
{\cal{R}}^{\not\epsilon}_{\not\gamma\not\alpha\not\beta}
&=&
\Gamma^{\not\epsilon}_{\not\alpha\not\beta\;,\not\gamma}
+\Gamma^{\not\epsilon}_{\not\alpha\not\gamma\;,\not\beta}+
\Gamma^{\not\rho}_{\not\gamma\not\alpha}\Gamma^{\not\epsilon}
_{\not\rho\not\beta}
+\Gamma^{\not\rho}_{\not\beta\not\alpha}
\Gamma^{\not\epsilon}_{\not\rho\not\gamma}
\nonumber\\
&\;&
+\Gamma^{\not\epsilon}_{\not\beta\not\gamma,\not\alpha}
+\Gamma^{\not\epsilon}_{\not\beta\not\alpha\;,\not\gamma}+
\Gamma^{\not\rho}_{\not\alpha\not\beta}\Gamma^{\not\epsilon}
_{\not\rho\not\gamma}
+\Gamma^{\not\rho}_{\not\gamma\not\beta}
\Gamma^{\not\epsilon}_{\not\rho\not\alpha}
\nonumber\\&\;&
+\Gamma^{\not\epsilon}_{\not\gamma\not\alpha\;,\not\beta}
+\Gamma^{\not\epsilon}_{\not\gamma\not\beta\;,\not\alpha}+
\Gamma^{\not\rho}_{\not\beta\not\gamma}\Gamma^{\not\epsilon}
_{\not\rho\not\alpha}
+\Gamma^{\not\rho}_{\not\alpha\not\gamma}
\Gamma^{\not\epsilon}_{\not\rho\not\beta}
\nonumber\\
&=&
\Gamma^{\not\epsilon}_{\not\alpha\not\beta\;,\not\gamma}
+\Gamma^{\not\epsilon}_{\not\alpha\not\gamma\;,\not\beta}+
\Gamma^{\not\rho}_{\not\gamma\not\alpha}\Gamma^{\not\epsilon}
_{\not\rho\not\beta}
+\Gamma^{\not\rho}_{\not\beta\not\alpha}
\Gamma^{\not\epsilon}_{\not\rho\not\gamma}
\nonumber\\
&\;&
-\Gamma^{\not\epsilon}_{\not\gamma\not\beta,\not\alpha}
-\Gamma^{\not\epsilon}_{\not\alpha\not\beta\;,\not\gamma}-
\Gamma^{\not\rho}_{\not\beta\not\alpha}\Gamma^{\not\epsilon}
_{\not\rho\not\gamma}
-\Gamma^{\not\rho}_{\not\beta\not\gamma}
\Gamma^{\not\epsilon}_{\not\rho\not\alpha}
\nonumber\\&\;&
-\Gamma^{\not\epsilon}_{\not\alpha\not\gamma\;,\not\beta}
+\Gamma^{\not\epsilon}_{\not\gamma\not\beta\;,\not\alpha}+
\Gamma^{\not\rho}_{\not\beta\not\gamma}\Gamma^{\not\epsilon}
_{\not\rho\not\alpha}
-\Gamma^{\not\rho}_{\not\gamma\not\alpha}
\Gamma^{\not\epsilon}_{\not\rho\not\beta}
\nonumber\\
&=&0
\end{eqnarray} and the same identity follows
immediately for the barred form. The
`curly-hat' form is defined as;
\begin{equation}
\hat{\cal{R}}^{\not\alpha}
_{\not\alpha\not\beta\not\gamma}=
\Gamma^{\not\rho}_{\not\gamma\not\alpha}\Gamma^{\not\alpha}
_{\not\rho\not\beta}
+\Gamma^{\not\rho}_{\not\beta\not\alpha}
\Gamma^{\not\alpha}_{\not\rho\not\gamma}
\end{equation}
and the following analogue of the
Bianci identity is proven for the 
$\hat{\cal{R}}_{\not\beta\not\gamma}
=\hat{\cal{R}}_{\not\gamma\not\beta}$
tensor;
\begin{eqnarray}
\hat{\cal{R}}_{\not\alpha\not\beta\;;\;\not\gamma}
+\hat{\cal{R}}_{\not\beta\not\gamma\;;\;\not\alpha}+
\hat{\cal{R}}_{\not\gamma\not\alpha\;;\;\not\beta}
=0
\label{cyclic}
\end{eqnarray}
To prove this identity we first need the identity;
\begin{equation}
u_{\not\alpha\;;\;\not\beta\;;\;\not\gamma\;;\;\not\delta}
+u_{\not\alpha\;;\;\not\beta\;;\;\not\delta\;;\;\not\gamma}
=
\bar{\cal{R}}^{\not\epsilon}_{\not\alpha\not\gamma\not\delta}
u_{\not\epsilon\;;\;\not\beta}
+
{\cal{R}}^{\not\epsilon}_{\not\beta\not\gamma\not\delta}
u_{\not\alpha\;;\;\not\epsilon}
\end{equation}
where;
\begin{equation}
{\bar{\cal{R}}}^{\not\phi}_{\not\alpha\not\beta\not\gamma}
=-\Gamma^{\not\phi}_{\not\alpha\not\beta\;,\not\gamma}
-\Gamma^{\not\phi}_{\not\alpha\not\gamma\;,\not\beta}+
\Gamma^{\not\rho}_{\not\gamma\not\alpha}\Gamma^{\not\phi}
_{\not\rho\not\beta}
+\Gamma^{\not\rho}_{\not\beta\not\alpha}
\Gamma^{\not\phi}_{\not\rho\not\gamma}
\end{equation}
the proof of which is lengthy and will not be reproduced here.
It then follows that;
\begin{eqnarray}
&(&u_{\not\alpha\;;\;\not\beta\;;\;\not\gamma\;;\;\not\delta}
+
u_{\not\alpha\;;\;\not\beta\;;\;\not\delta\;;\;\not\gamma})
+
(u_{\not\alpha\;;\;\not\delta\;;\;\not\beta\;;\;\not\gamma}
+
u_{\not\alpha\;;\;\not\delta\;;\;\not\gamma\;;\;\not\beta})
+
(u_{\not\alpha\;;\;\not\gamma\;;\;\not\delta\;;\;\not\beta}
+
u_{\not\alpha\;;\;\not\gamma\;;\;\not\beta\;;\;\not\delta})
\nonumber\\
&=&
\bar{\cal{R}}^{\not\epsilon}_{\not\alpha\not\gamma\not\delta}
u_{\not\epsilon\;;\;\not\beta}
+
{\cal{R}}^{\not\epsilon}_{\not\beta\not\gamma\not\delta}
u_{\not\alpha\;;\;\not\epsilon}
+
\bar{\cal{R}}^{\not\epsilon}_{\not\alpha\not\beta\not\gamma}
u_{\not\epsilon\;;\;\not\delta}
+
{\cal{R}}^{\not\epsilon}_{\not\delta\not\beta\not\gamma}
u_{\not\alpha\;;\;\not\epsilon}
+
\bar{\cal{R}}^{\not\epsilon}_{\not\alpha\not\delta\not\beta}
u_{\not\epsilon\;;\;\not\gamma}
+
{\cal{R}}^{\not\epsilon}_{\not\gamma\not\delta\not\beta}
u_{\not\alpha\;;\;\not\epsilon}
\nonumber\\
&=&
\bar{\cal{R}}^{\not\epsilon}_{\not\alpha\not\gamma\not\delta}
u_{\not\epsilon\;;\;\not\beta}
+
\bar{\cal{R}}^{\not\epsilon}_{\not\alpha\not\beta\not\gamma}
u_{\not\epsilon\;;\;\not\delta}
+
\bar{\cal{R}}^{\not\epsilon}_{\not\alpha\not\delta\not\beta}
u_{\not\epsilon\;;\;\not\gamma}
\label{bianci}
\end{eqnarray}
viz eq.(\ref{cyclic}) 
whilst the first line  of eq.(\ref{bianci}) can alternatively
be written as;
\begin{eqnarray}
&(&\;{\cal{R}}^{\not\epsilon}_{\not\alpha\not\beta\not\gamma}
u_{\not\epsilon}\;)_{\;;\;\not\delta}
+
(\;{\cal{R}}^{\not\epsilon}_{\not\alpha\not\gamma\not\delta}
u_{\not\epsilon}\;)_{\;;\;\not\beta}
+
(\;{\cal{R}}^{\not\epsilon}_{\not\alpha\not\delta\not\beta}
u_{\not\epsilon}\;)_{\;;\;\not\gamma}
\nonumber\\
&=&
-({\cal{R}}^{\not\epsilon}_{\not\alpha\not\beta
\not\gamma\;;\;\not\delta}+
{\cal{R}}^{\not\epsilon}_{\not\alpha\not\gamma
\not\delta\;;\;\not\beta}+
{\cal{R}}^{\not\epsilon}_{\not\alpha\not\delta
\not\beta\;;\;\not\gamma})\;u_{\not\epsilon}
+{\cal{R}}^{\not\epsilon}_{\not\alpha\not\beta\not\gamma}
u_{\not\epsilon\;;\;\not\delta}
+{\cal{R}}^{\not\epsilon}_{\not\alpha\not\gamma\not\delta}
u_{\not\epsilon\;;\;\not\beta}
+{\cal{R}}^{\not\epsilon}_{\not\alpha\not\delta\not\beta}
u_{\not\epsilon\;;\;\not\gamma}
\label{bianci2}
\end{eqnarray}
so we must have;
\begin{equation}
{\cal{R}}^{\not\epsilon}_{\not\alpha\not\beta
\not\gamma\;;\;\not\delta}+
{\cal{R}}^{\not\epsilon}_{\not\alpha\not\gamma
\not\delta\;;\;\not\beta}+
{\cal{R}}^{\not\epsilon}_{\not\alpha\not\delta
\not\beta\;;\;\not\gamma}=
({{\cal{R}}}^{\not\epsilon}_{\not\alpha\not\beta\not\gamma}
-
{\bar{\cal{R}}}^{\not\epsilon}_{\not\alpha\not\beta\not\gamma})
u_{\not\epsilon\;;\;\not\delta}+
\mbox{plus cyclic permutations.}
\label{rot}
\end{equation}
Now since the connection is antisymmetric in its lower
two indices when the derivative of the symmetric part of the
metric vanishes (the case under consideration here);
\begin{eqnarray}
\Gamma^{\not\alpha}_{\not\alpha\not\beta}
&=&
{1\over2}(g_{\not\epsilon\not\alpha\,,\,\not\beta}
+
g_{\not\beta\not\epsilon\,,\,\not\alpha}
-
g_{\not\alpha\not\beta\,,\,\not\epsilon})
\bar{g}^{\not\epsilon\not\alpha}
\\
\nonumber
&=&
{1\over2}(g^{\;\;\not\alpha}_{\not\epsilon\;\;\;,\,\not\beta}
+
g_{\not\beta\not\epsilon\,,\,}^{\;\;\;\;\;\not\alpha}
-
g^{\not\alpha}_{\;\;\not\beta\,,\,\not\epsilon})
\bar{g}^{\not\epsilon}_{\;\;\not\alpha}
\\
\nonumber
&=&
{-1\over2}(g^{\;\;\not\alpha}_{\not\epsilon\;\;\;,\,\not\alpha}
+
g_{\not\alpha\not\epsilon\,,\,}^{\;\;\;\;\;\not\alpha}
-
g^{\not\alpha}_{\;\;\not\alpha\,,\,\not\epsilon})
\bar{g}^{\not\epsilon}_{\;\;\not\beta}
\\
\nonumber&=&0
\end{eqnarray}
so that;
\begin{equation}
{{\cal{R}}}^{\not\alpha}_{\not\alpha\not\beta\not\gamma}
-
{\bar{\cal{R}}}^{\not\alpha}_{\not\alpha\not\beta\not\gamma}
=
2\Gamma^{\not\alpha}_{\not\alpha\not\gamma\,,\,\not\beta}
+2\Gamma^{\not\alpha}_{\not\alpha\not\beta\,,\,\not\gamma}
=0
\end{equation}
and hence  we finally obtain from eq.(\ref{rot});
\begin{equation}
{\hat{\cal{R}}}_{\not\beta
\not\gamma\;;\;\not\delta}+
{\hat{\cal{R}}}_{\not\gamma
\not\delta\;;\;\not\beta}+
{\hat{\cal{R}}}_{\not\delta
\not\beta\;;\;\not\gamma}=0
\end{equation}
because the $u_{\alpha}$ is arbitrary.
The minus sign in the last line of eq.(\ref{bianci2})
arises because the vector field in question, 
here $u_{\not\!\alpha}$,
must satisfy the L\"{o}rentz condition. This follows
automatically since for the anti-symmetric part
of the metric which contains the potential
of the electro-magnetic field and has non-vanishing 
derivatives;
\begin{equation}
\Gamma^{\not\alpha}_{\not\alpha\not\beta}
=0\;\;\mbox{iff}\;\;g^{a\;\;\;,\,\not\alpha}
_{\not\alpha\not\beta}=0
\Rightarrow
\partial_{\alpha}A^{\alpha}=0
\label{Lorentz2}
\end{equation}
because the arbitrary $\beta$ index will be eliminated
in the gauge invariant terms which appear in the
stress-energy tensor,
so that commuting the derivative index $\not\!\!\beta$ 
 past the vector
index $\not\!\!\alpha$ generates a minus sign plus a 
$2\eta_{\alpha\beta}
\partial^{\not\phi}u_{\not\phi}=0$ i.e. 
anti-commutes past the vector index).

\end{document}